\documentclass{mn2e}
\usepackage{hyperref}
\usepackage{graphicx}

\usepackage{natbib}
\usepackage{astroj}
\usepackage{times}

\usepackage{times} \usepackage{color}

\title[ NGC 3680: element abundances]{Elemental abundances of intermediate age open cluster NGC 3680 \thanks{Based on observations obtained at the European Southern observatory, Paranal, Chile (ESO programme 072.B-0331(B)).}}
\author[ Mitschang \& De Silva \& Zucker]{A. W. Mitschang $^{1,2}$ \thanks{ E-mail: arik.mitschang@mq.edu.au}, G. M. De Silva $^{3}$, D. B. Zucker $^{1,2,3}$\\
$^1$ Macquarie University Research Centre in Astronomy, Astrophysics \& Astrophotonics \\
$^2$ Department of Physics \& Astronomy, Macquarie University, NSW 2109 Australia \\
$^3$ Australian Astronomical Observatory, PO Box 296, Epping NSW 1710 }
\date{\today}

\begin{document}

\maketitle

\begin{abstract}
\label{sec-1}

We present a new abundance analysis of the intermediate age Galactic
open cluster NGC 3680, based on high resolution, high signal-to-noise
VLT/UVES spectroscopic data. Several element abundances are presented
for this cluster for the first time, but most notably we derive
abundances for the light and heavy s-process elements Y, Ba, La, and
Nd. The serendipitous measurement of the rare-earth r-process element
Gd is also reported. This cluster exhibits a significant enhancement
of Na in giants as compared to dwarfs, which may be a proxy for an O
to Na anti-correlation as observed in Galactic globular clusters but
not open clusters. We also observe a step-like enhancement of heavy
s-process elements towards higher atomic number, contrary to
expectations from AGB nucleosynthesis models, suggesting that the
r-process played a significant role in the generation of both La and
Nd in this cluster.
\end{abstract}

\section{Introduction}
\label{sec-2}

Galactic open clusters (OCs) have long been established as important
fossils of both the dynamical and chemical evolution of the
Galaxy. They are numerous, enable reasonably accurate dating, and even
though their lifetimes are typically thought to be very short
\citep{janes1988}, significant numbers of old OCs have been detected
ranging back to the primordial Galaxy
\citep{phelps1994,liu2000,randich2009,bragaglia2008,sestito2008}. Hence
they provide a time-resolved sample, across a range of Galacto-centric
radii, with which to probe the chemical evolution of the Galaxy
\citep{friel2002}. Typically, [Fe/H] abundances are used as tracers of
chemical evolution across OCs, and there are a large number of
clusters for which these values have been computed, but only a modest
sample of studies go beyond this basic indicator (e.g., see
\citealt{carrera2011}), into the realm of the $\alpha$, r-, and s-process
elements.

These elements are now recognized to be important generational
indicators, and form a basis by which to piece together the primordial
building blocks of the Galaxy \citep{fbh2002}. It has been demonstrated
that their patterns, which become imprinted in the photospheres of
cluster members when they are born, can be used to link dissolved
associations to their natal environments
\citep{desilva2006,desilva2007,bubar2010}.

The nucleosynthesis of these elements is believed to happen in the
very late phases of stellar evolution \citep{wallerstein1997};
specifically, Asymptotic Giant Branch (AGB) phase stars are thought to
be responsible for the enrichment of some s-process elements including
Y, Ba, Nd, and La, while high mass stars ($M\ge{}8 M_{\odot}$) that are
fated for death via core-collapse supernova (SN II) are responsible
for a range of the $\alpha${}-elements including Mg, Si, Ca, and
Ti. Type Ia supernovae (SN Ia) generate large amounts of the Iron-peak
elements including Fe, Ni, Cr, Co and Mn.  Several authors have
established trends for these elements amongst thin and thick Galactic
disk field stars for a range of metallicities and ages
\citep{bensby2005,bensby2003,reddy2003}, enabling them to make
conjectures about the relative rates and timescales at which these
types of events occurred.

NGC 3680, an intermediate age ($\sim${}1.5 Gyr) OC situated about 8 kpc
from the Galactic center in the thin disk, has been extensively
studied photometrically and to some extent spectroscopically.
\citet{nordstroem1997} established membership probabilities based on
radial velocities (RVs) and proper motions, and though there has been
some spread in the literature over cluster mean [Fe/H] values, from
$\sim${}$-$0.17 dex to $\sim${}0.09 dex
\citep{nordstroem1997,bruntt1999,pasquini2001,anthony-twarog2009}, the
general consensus from recent studies is that this cluster is slightly
metal poor at $\sim${}$-$0.08 dex. Most spectroscopic studies have been
targeted at understanding the nature of the Li-dip observed in this
and other clusters \citep{anthony-twarog2009,pasquini2001}, and thus
there is a lack of published abundances beyond Li and Fe, though
\citet{anthony-twarog2009} derived Si, Ca, and Ni abundances for their
program stars, and \citet{pace2008} reported several metal abundances for
two dwarf stars.

In this paper we present an abundance analysis of 8 giants and 3
dwarfs based on high resolution spectroscopic data, covering 14
elements, including several s-process and one rare-earth r-process
element, Gadolinium. In Section \hyperref[sec-3]{\ref{sec-3}} we describe
our program targets and observations and describe the methods used in
determining differential abundances and uncertainties. In Section
\hyperref[sec-4]{\ref{sec-4}} we discuss membership, general characteristics and
abundance trends of NGC 3680, and finally in Section \hyperref[sec-5]{\ref{sec-5}} we
summarise our findings.
\section{Observations \& Analysis}
\label{sec-3}

The targets in this study were selected from the catalog of
\citet{nordstroem1997}, as members of NGC 3680, using the membership
determination criteria therein. Observations were carried out using
the FLAMES multi-object fibre system feeding the UVES spectrograph
(R$\sim${}45,000) at the VLT as a part of service program 072.B-0331B
over 9 nights in 2004. Table \ref{targets} gives details of the targets
in the study, including stellar parameters as determined below, where
the top part shows giants and the bottom part dwarfs. Individual
objects were observed at several distinct times; all spectroscopic
analyses herein were performed on exposure-weighted averages of all
available spectra for each object. Care was taken to ensure matching
of the wavelength scale for the different epochs observed before
combining. Typical signal-to-noise (S/N) was $\sim${}130 for the giants
and $\sim${}90 for the dwarfs (star Nos. 5, 24, 35).

\label{ref-targets}
\begin{table}
\caption{Target data} \label{targets}
\begin{tabular}{rrrrrrr}
\hline
\hline
 No.  &          RA  &          Dec  &  V$_{\mathrm{mag}}$  &  T$_{\mathrm{eff}}$  &            log \emph{g}  &   $\xi$  \\
      &     (h:m:s)  &      (d:m:s)  &                      &                 (K)  &  (cm s$^{\mathrm{-2}}$)  &  (km/s)  \\
\hline
  11  &  11:25:29.2  &  $-$43:15:48.0  &               10.88  &                5100  &                     3.1  &    1.60  \\
  13  &  11:25:16.1  &  $-$43:14:24.3  &               10.78  &                4950  &                     3.0  &    1.69  \\
  20  &  11:25:26.2  &  $-$43:11:24.2  &               10.10  &                5200  &                     3.2  &    1.61  \\
  26  &  11:25:38.0  &  $-$43:16:06.4  &               10.92  &                5100  &                     3.3  &    1.73  \\
  27  &  11:25:41.9  &  $-$43:17:07.0  &               10.73  &                5050  &                     3.1  &    1.78  \\
  34  &  11:25:38.6  &  $-$43:13:58.9  &               10.60  &                5100  &                     2.9  &    0.94  \\
  41  &  11:25:48.4  &  $-$43:09:52.7  &               10.88  &                5100  &                     3.2  &    1.76  \\
  44  &  11:25:49.8  &  $-$43:12:16.0  &                9.98  &                4800  &                     2.6  &    2.10  \\
\hline
  24  &  11:25:34.3  &  $-$43:15:21.6  &               13.77  &                6650  &                     4.8  &    2.13  \\
  35  &  11:25:38.1  &  $-$43:13:27.3  &               13.07  &                6800  &                     4.4  &    2.20  \\
   5  &  11:25:18.4  &  $-$43:16:25.0  &               12.83  &                7100  &                     4.1  &    2.20  \\
\hline
\end{tabular}
\end{table}
\subsection{Equivalent width measurements}
\label{sec-3-1}

\begin{figure}
\centering
\includegraphics[angle= 270,width=\linewidth]{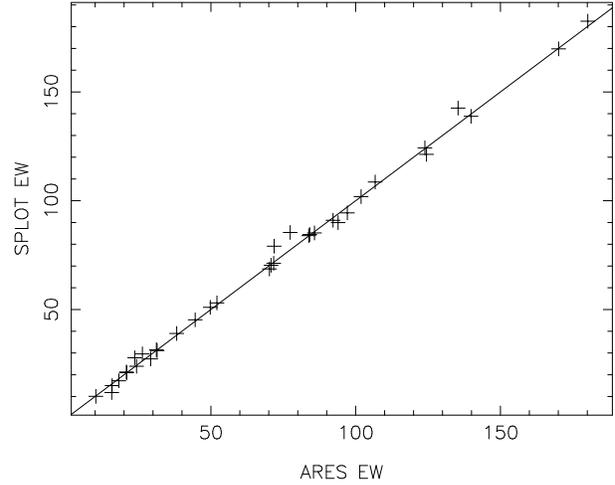}
\caption{\label{arescompare}Comparison between ARES and hand measured EWs for a random sample of lines. The solid line indicates a one to one relationship.}
\end{figure}

Absorption line equivalent widths (EWs) were measured using the
ARES\footnote{http://www.astro.up.pt/\~{}sousasag/ares/ } code
\citep{aresref} for automatic equivalent width
measurements. Briefly, the code works by first determining the
continuum level using an iterative point rejection technique,
normalising the data, locating separate absorption features by looking
for crossings in the third differential of the data and then fitting
Gaussian profiles to these features. There are several input
parameters, the optimal values of which were determined by comparing
hand measurements of EWs in a typical spectrum using the \emph{splot}
function in IRAF\footnote{http://iraf.noao.edu/ } until the best agreement
was achieved. Figure \ref{arescompare} illustrates the performance of
the ARES code on these high resolution data. The results of EW
measurement were further checked by reviewing plots of the continuum
and Gaussian profile fits for each line in each target. Where there
were anomalous measurements (e.g. continuum contaminated by a spurious
cosmic ray) the ARES measurements were replaced by hand measurements
of those features. Table \ref{ewtable} lists the atomic parameters
$\lambda$, EP, and log \emph{gf} for each species along with measured EWs for
each star.

\begin{table*}
\caption{Equivalent Widths for program stars} \label{ewtable}
\begin{tabular}{rlrrrrrrrrrrrrr}
\hline
\hline
 $\lambda$  &  ion   &    EP  &  log \emph{gf}  &      11  &      13  &      20  &      26  &      27  &      34  &     41  &      44  &     24  &     35  &      5  \\
     \AA{}  &        &        &                 &          &          &          &          &          &  m\AA{}  &         &          &         &         &         \\
\hline
   4799.41  &  Fe I  &  3.64  &          $-$2.23  &   68.25  &   75.40  &   67.43  &   74.30  &   73.38  &   49.78  &  74.28  &   85.48  &  19.58  &  12.63  &   7.30  \\
   4802.88  &  Fe I  &  3.64  &          $-$1.51  &   88.31  &   92.50  &   88.08  &   91.32  &   92.29  &   64.86  &  93.29  &   94.64  &  47.72  &  40.99  &  28.40  \\
   4808.15  &  Fe I  &  3.25  &          $-$2.79  &   62.40  &   70.62  &   58.60  &   66.12  &   68.93  &   49.96  &  62.75  &   73.66  &  15.20  &         &   4.51  \\
   4809.94  &  Fe I  &  3.57  &          $-$2.72  &   45.76  &   52.83  &   45.26  &   53.19  &   50.75  &   39.79  &  48.97  &   57.12  &  10.84  &         &         \\
   4835.87  &  Fe I  &  4.10  &          $-$1.50  &  102.19  &  113.20  &  100.48  &  113.92  &  111.55  &   83.96  &  59.23  &  102.64  &  36.46  &  25.23  &  19.81  \\
\hline
\end{tabular}
\end{table*}
\subsection{Stellar parameter determinations}
\label{sec-3-2}

The stellar parameters log \emph{g}, T$_{\mathrm{eff}}$, and microturbulent velocity $\xi$,
form the basis required to translate EWs to abundances or to fit
synthetic spectra, based on assumed abundances, to observed
spectra. We determined these parameters using spectroscopic methods
and a grid of ATLAS9 \citep{atlas9ref} model atmospheres interpolated
to give a resolution of 50K in T$_{\mathrm{eff}}$ and 0.1 in log \emph{g}. Using the
MOOG\footnote{http://www.as.utexas.edu/\~{}chris/moog.html } \citep{moogref}
driver \emph{abfind} to force-fit abundances to our measured EWs, stellar
parameters were determined by hand for each star in the typical way:
T$_{\mathrm{eff}}$ by balancing \mbox{Fe\,{\sc i}} abundances against excitation
potential, $\xi$ by balancing \mbox{Fe\,{\sc i}} abundances against
reduced EW, and finally log \emph{g} by requiring ionization balance
between \mbox{Fe\,{\sc i}} and \mbox{Fe\,{\sc ii}}. On average across
our eleven targets 120 \mbox{Fe\,{\sc i}} lines and 11 \mbox{Fe\,{\sc
ii}} lines were used in the determination of stellar parameters.
\subsection{Abundance Analysis}
\label{sec-3-3}

The atomic line data for Fe and the other EW species are a subset of those
used by \citet{bensby2003}, Ba data were taken from \citet{mcwilliam1998}
while La, Nd and Gd were obtained via the VALD
database\footnote{http://vald.astro.univie.ac.at/\~{}vald/ } \citep{valdref}.
Abundances for the elements, Fe, Zn, Ca, Na, Mg, Cr, Ni, Ti, Si, and
Al were then force-fit using \emph{abfind} to all measured EWs for a given
species. Values relative to solar (i.e., [X/Fe], solar values derived
as described below) are shown in Table \ref{abundances}, along with
separate cluster means for giants (top) and dwarfs (bottom).

\begin{figure}
\centering
\includegraphics[width=\linewidth]{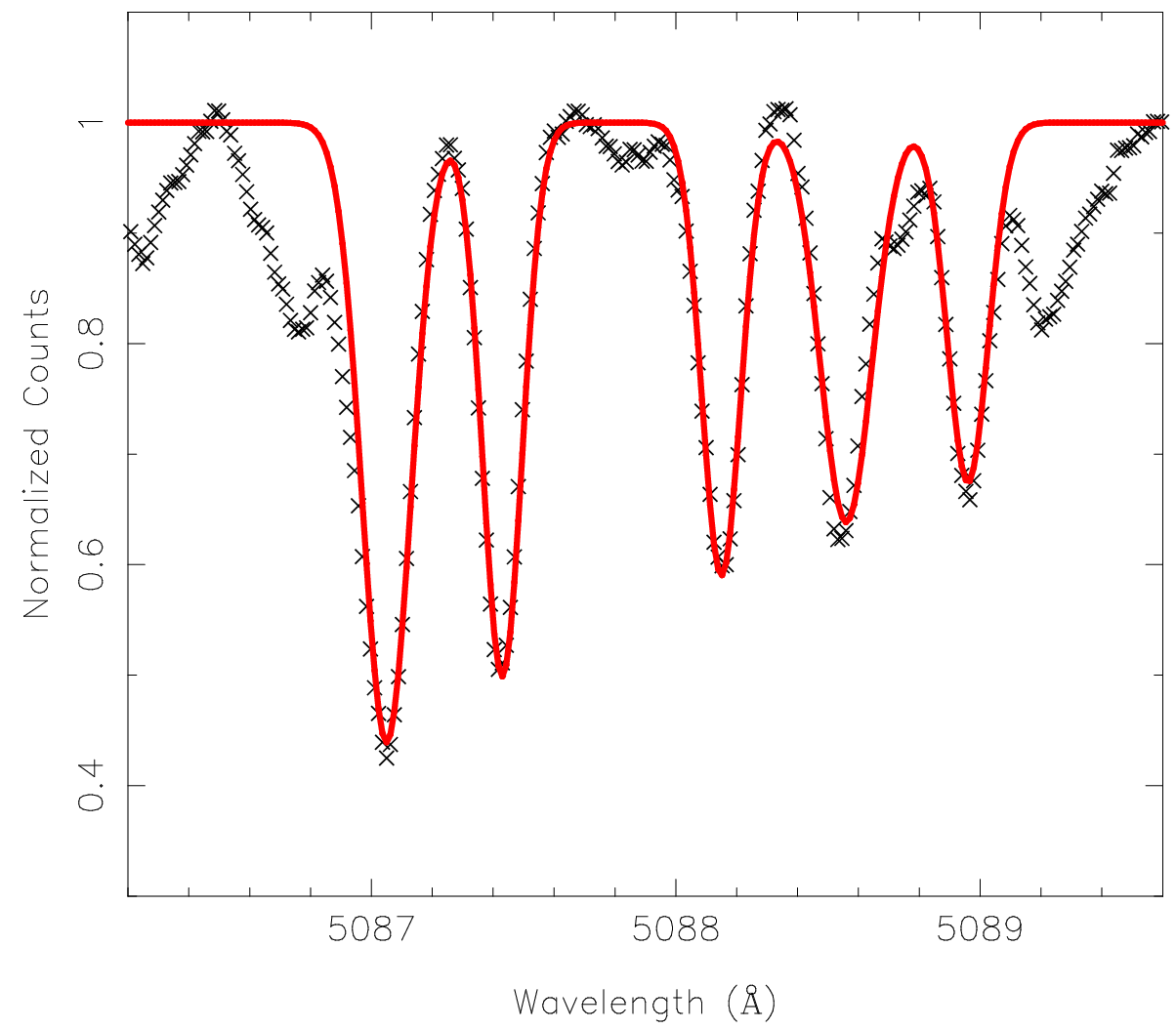}
\caption{\label{linefit}ARES automatic line fitting for a small region of one of these high-resolution spectra. Only lines used in our analysis are shown here, but the code fits all features in the region for accuracy, including the features to the left and right of the line group shown here.}
\end{figure}

For the species Ba, La, Nd, Gd, whose lines were too blended or for
which the effects of hyperfine splitting preclude simple EW
measurements, synthetic spectra were generated and compared to
observations using the MOOG driver \emph{synth}. A range of abundance
values were selected and compared by eye until a suitable fit was
identified. Spectrum synthesis was not performed on the three dwarfs
because of the lower S/N. The results of synthesis are reported
together with other abundances in Table \ref{abundances}.

Recent work by \citet{santos2009} derived spectral parameters for 3
giants in common with those in this study using VLT UVES data at
R$\sim${}50,000. The T$_{\mathrm{eff}}$, log \emph{g}, and $\xi$ values obtained from our
analysis are systematically higher by, on average, 250K, 0.4, and 0.3
km/s, respectively. To address any possible systematics in terms of
atomic parameters, we repeated our stellar parameter determination
using Fe atomic data from the same line list \citep{sousa2008} as in
\citet{santos2009}. Given the differing spectral range in our data, we
were only able to utilize 181/263 \mbox{Fe\,{\sc i}} lines and 24/36
\mbox{Fe\,{\sc ii}} lines. With this analysis we came out with
parameters compatible with our previous measurements. To further test
our Fe atomic parameters, in the same manner as in \citet{santos2009}, we
derived stellar parameters for the HARPS solar spectrum ``Ganymede''
again with the same procedure followed above, obtaining very close to
expected values (T$_{\mathrm{eff}}$$=$5850$\pm${}50, log \emph{g}$=$4.4$\pm${}0.1,
$\xi$$=$0.97$\pm${}0.1, [Fe/H]$=$0.02$\pm${}0.08). Therefore, the differences in
stellar parameters are likely due to differences in data and code used
to derive the them.

Systematic errors affecting one's ability to compare abundance results
from different studies come from a variety of sources, including
choices of model atmosphere grids, atomic data and more subtle effects
such as automated algorithms for continuum determination. In order to
minimise the contribution of such systematics, all abundance values
presented here are reported relative compared to solar measurements as
described above, using the ``Ganymede'' spectrum. We derive solar values
via EWs for Fe, Na, Mg, Al, Si, Ca, Ti, Cr, Ni, Zn, and Y of 7.54,
6.43, 7.79, 6.52, 7.53, 6.42, 4.92, 5.73, 6.28, 4.69, and 2.17,
respectively. For the synthesised elements Ba, La, and Nd, our
computed values are 2.23, 1.12, and 1.45.
\subsection{Errors and Uncertainties}
\label{sec-3-4}

The sources of uncertainty in abundance values are related to the
measurement of EWs (including the automatic continuum determination),
the atomic parameters (including excitation potential and log \emph{gf}),
and the determination of stellar parameters (which in turn are derived
from Fe EW measurements, atomic parameters and choice of atmospheric
model).

Uncertainties for EWs were estimated via a Monte-Carlo method, by
inducing variations in the most sensitive ARES parameter, \emph{rejt} (as
noted by \citealt{aresref}), between reasonable values, (0.990,0.999) for
giants and (0.885,0.995) for dwarfs (the difference in mean \emph{rejt}
values between dwarfs and giants being a function of S/N). We collated
measurement differences from the mean across all lines in a given
species for all stars. The resulting distribution was approximately
Gaussian with the zero point being the EW used in analysis; the 90\%
limit of this distribution (i.e., 90\% of all differences lie within
this value) was then used in determining the abundance sensitivity for
each species for a representative giant and dwarf.

Uncertainties were conservatively assumed to be at the measurement
resolution for T$_{\mathrm{eff}}$ and log \emph{g}, 50K and 0.1 respectively, and the
uncertainty in $\xi$ was taken to be 0.1 km/s. Abundance sensitivities
are reported in Table \ref{uncertainties} for all identified sources of
error, where $\delta${}L and $\delta${}T are the uncertainty on log \emph{g}
and T$_{\mathrm{eff}}$ respectively. The quadrature sum of all these sources is
reported as the total uncertainty. Due to the marked difference in
quality of spectra for our giants and dwarfs, we have computed the
uncertainties for each separately.

\begin{table*}
\caption{Abundance uncertainty sensitivities} \label{uncertainties}
\begin{tabular}{llllllllllllllll}
\hline
                            &  Fe           &  Na           &  Mg           &  Al           &  Si           &  Ca           &  Ti           &  Cr           &  Ni           &  Zn           &  Y            &  Ba           &  La           &  Nd           &  Gd           \\
\hline
 $\delta${}L$\pm${}0.1      &  $\pm${}0.01  &  $\pm${}0.05  &  $\pm${}0.01  &  $\pm${}0.01  &  $\pm${}0.02  &  $\mp${}0.01  &  $\pm${}0.02  &  $\pm${}0.01  &  $\pm${}0.01  &  $\pm${}0.03  &  $\mp${}0.01  &  $\pm${}0.05  &  $\pm${}0.04  &  $\pm${}0.04  &  $\pm${}0.04  \\
 $\delta${}T$\pm${}50       &  $\pm${}0.04  &  $\mp${}0.04  &  $\pm${}0.03  &  $\pm${}0.03  &  $\mp${}0.01  &  $\pm${}0.04  &  $\pm${}0.04  &  $\pm${}0.04  &  $\pm${}0.03  &  $\mp${}0.01  &  $\pm${}0.06  &  $\pm${}0.01  &  $\pm${}0.01  &  $\pm${}0.01  &  $\mp${}0.01  \\
 $\delta${}$\xi$$\pm${}0.1  &  $\mp${}0.04  &  $\mp${}0.03  &  $\mp${}0.03  &  $\pm${}0.01  &  $\mp${}0.01  &  $\mp${}0.04  &  $\pm${}0.03  &  $\mp${}0.02  &  $\mp${}0.04  &  $\mp${}0.06  &  $\mp${}0.07  &  $\mp${}0.03  &  $\mp${}0.02  &  $\mp${}0.04  &  $\mp${}0.03  \\
 $\delta${}EW               &  $\pm${}0.06  &  $\pm${}0.09  &  $\pm${}0.06  &  $\pm${}0.06  &  $\pm${}0.07  &  $\pm${}0.07  &  $\pm${}0.03  &  $\pm${}0.05  &  $\pm${}0.06  &  $\pm${}0.00  &  $\pm${}0.03  &  $\pm${}0.04  &  $\pm${}0.04  &  $\pm${}0.06  &  $\pm${}0.06  \\
\hline
 Tot$_{\mathrm{giant}}$     &  $\pm${}0.08  &  $\pm${}0.11  &  $\pm${}0.07  &  $\pm${}0.07  &  $\pm${}0.08  &  $\pm${}0.09  &  $\pm${}0.06  &  $\pm${}0.07  &  $\pm${}0.08  &  $\pm${}0.06  &  $\pm${}0.10  &  $\pm${}0.07  &  $\pm${}0.06  &  $\pm${}0.08  &  $\pm${}0.08  \\
\hline
 $\delta${}L$\pm${}0.1      &  $\pm${}0.00  &  $\pm${}0.01  &  $\mp${}0.01  &               &  $\pm${}0.00  &  $\mp${}0.01  &  $\pm${}0.00  &  $\pm${}0.00  &  $\pm${}0.00  &  $\pm${}0.01  &  $\pm${}0.01  &               &               &               &               \\
 $\delta${}T$\pm${}50       &  $\pm${}0.03  &  $\pm${}0.02  &  $\pm${}0.02  &               &  $\pm${}0.01  &  $\pm${}0.03  &  $\pm${}0.04  &  $\pm${}0.03  &  $\pm${}0.03  &  $\pm${}0.03  &  $\pm${}0.04  &               &               &               &               \\
 $\delta${}$\xi$$\pm${}0.1  &  $\mp${}0.01  &  $\pm${}0.01  &  $\mp${}0.01  &               &  $\mp${}0.01  &  $\mp${}0.01  &  $\mp${}0.01  &  $\pm${}0.00  &  $\mp${}0.01  &  $\mp${}0.01  &  $\mp${}0.01  &               &               &               &               \\
 $\delta${}EW               &  $\pm${}0.07  &  $\pm${}0.12  &  $\pm${}0.04  &               &  $\pm${}0.07  &  $\pm${}0.06  &  $\pm${}0.06  &  $\pm${}0.16  &  $\pm${}0.07  &  $\pm${}0.00  &  $\pm${}0.01  &               &               &               &               \\
\hline
 Tot$_{\mathrm{dwarf}}$     &  $\pm${}0.07  &  $\pm${}0.12  &  $\pm${}0.05  &               &  $\pm${}0.08  &  $\pm${}0.06  &  $\pm${}0.07  &  $\pm${}0.17  &  $\pm${}0.07  &  $\pm${}0.03  &  $\pm${}0.04  &               &               &               &               \\
\hline
\end{tabular}
\end{table*}

\begin{table*}
\caption{Differential [X/Fe] abundances} \label{abundances}
\begin{tabular}{rrrrrrrrrrrrrrrr}
\hline
  No.  &  [Fe/H]  &     Na  &     Mg  &     Al  &     Si  &     Ca  &    Ti  &     Cr  &     Ni  &     Zn  &     Y  &    Ba  &    La  &    Nd  &     Gd  \\
\hline
   11  &    0.04  &   0.05  &  $-$0.05  &  $-$0.06  &   0.04  &  $-$0.15  &  0.18  &   0.01  &  $-$0.07  &  $-$0.34  &  0.25  &  0.16  &  0.36  &  0.56  &  $-$0.39  \\
   13  &   $-$0.01  &   0.13  &   0.00  &   0.12  &   0.07  &  $-$0.09  &  0.22  &   0.09  &   0.01  &  $-$0.38  &  0.23  &  0.16  &  0.31  &  0.56  &  $-$0.34  \\
   20  &    0.06  &   0.22  &  $-$0.14  &   0.08  &   0.10  &  $-$0.09  &  0.24  &   0.04  &  $-$0.05  &  $-$0.31  &  0.10  &  0.18  &  0.24  &  0.44  &  $-$0.46  \\
   26  &    0.09  &   0.12  &  $-$0.01  &   0.16  &   0.01  &  $-$0.03  &  0.29  &   0.11  &  $-$0.03  &  $-$0.43  &  0.13  &  0.12  &  0.29  &  0.51  &  $-$0.49  \\
   27  &    0.04  &   0.16  &  $-$0.01  &   0.10  &  $-$0.02  &  $-$0.11  &  0.21  &   0.10  &  $-$0.10  &  $-$0.50  &  0.09  &  0.06  &  0.26  &  0.56  &  $-$0.34  \\
   34  &   $-$0.11  &   0.04  &  $-$0.15  &   0.02  &  $-$0.03  &  $-$0.09  &  0.07  &   0.07  &  $-$0.04  &  $-$0.48  &  0.19  &  0.21  &        &  0.21  &  $-$0.64  \\
   41  &    0.03  &   0.11  &  $-$0.01  &   0.18  &   0.03  &  $-$0.06  &  0.25  &   0.12  &   0.00  &  $-$0.42  &  0.11  &  0.13  &  0.27  &  0.55  &  $-$0.43  \\
   44  &   $-$0.13  &   0.18  &   0.05  &   0.27  &   0.15  &  $-$0.03  &  0.21  &   0.14  &  $-$0.03  &  $-$0.49  &  0.24  &  0.03  &  0.43  &  1.09  &  $-$0.49  \\
\hline
 Mean  &    0.00  &   0.13  &  $-$0.04  &   0.11  &   0.04  &  $-$0.08  &  0.21  &   0.09  &  $-$0.04  &  $-$0.42  &  0.17  &  0.13  &  0.31  &  0.56  &  $-$0.45  \\
\hline
   24  &   $-$0.07  &  $-$0.16  &  $-$0.16  &         &   0.00  &  $-$0.05  &  0.15  &   0.04  &  $-$0.03  &  $-$0.29  &  0.11  &        &        &        &         \\
   35  &   $-$0.15  &  $-$0.06  &  $-$0.11  &         &   0.01  &  $-$0.04  &  0.07  &  $-$0.07  &  $-$0.10  &  $-$0.23  &  0.12  &        &        &        &         \\
    5  &   $-$0.17  &  $-$0.15  &  $-$0.11  &         &   0.01  &  $-$0.01  &  0.11  &  $-$0.16  &  $-$0.09  &  $-$0.16  &  0.11  &        &        &        &         \\
\hline
 Mean  &   $-$0.16  &  $-$0.10  &  $-$0.11  &         &   0.01  &  $-$0.03  &  0.09  &  $-$0.12  &  $-$0.10  &  $-$0.20  &  0.12  &        &        &        &         \\
\hline
\end{tabular}
\end{table*}
\section{Discussion}
\label{sec-4}
\subsection{Cluster membership and properties}
\label{sec-4-1}

\begin{figure}
\centering
\includegraphics[width=\linewidth]{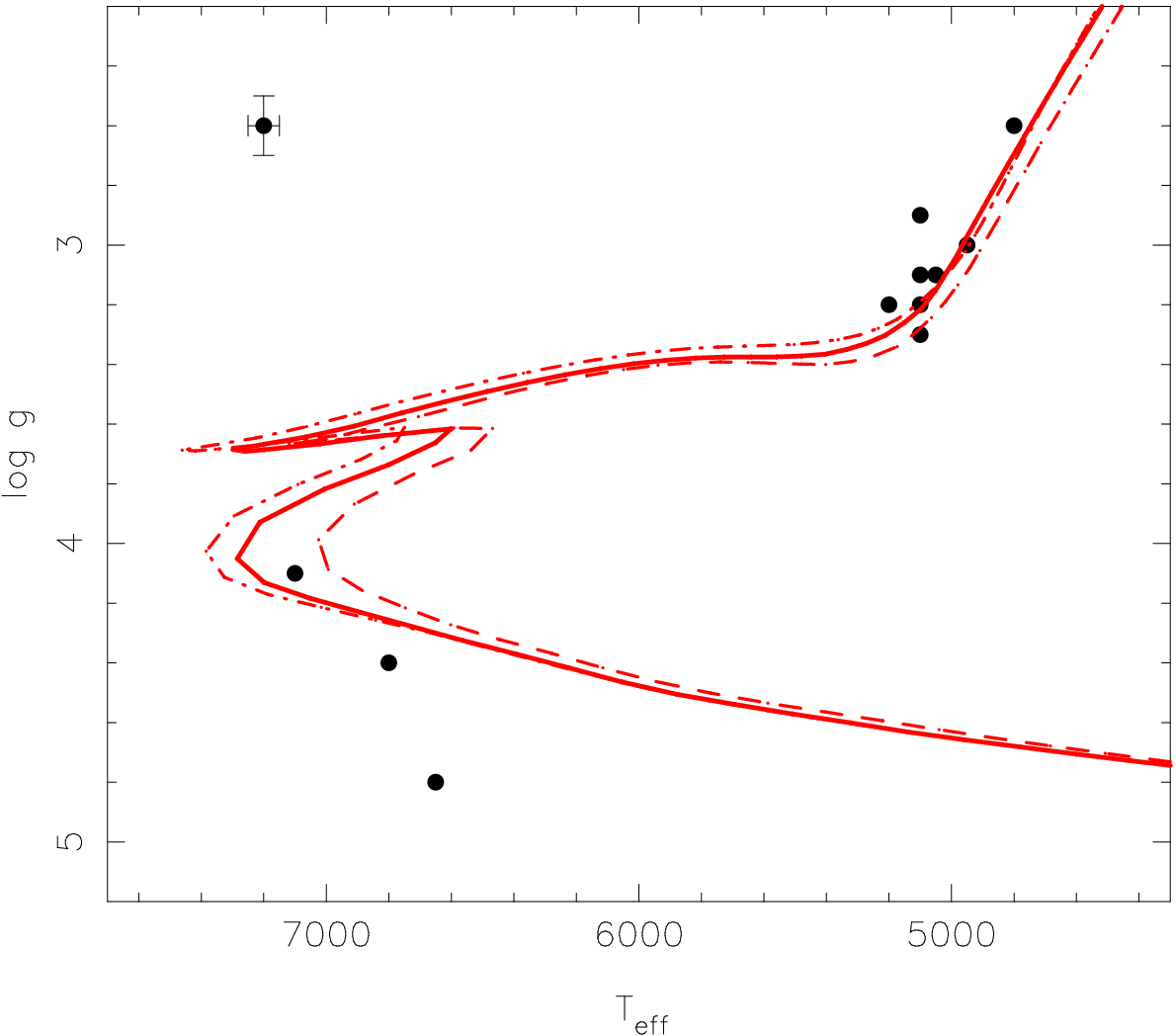}
\caption{\label{yyfit}Best fit Y$^2$ isochrone to derived stellar parameters. The solid line represents the best fit isochrone of age 1.4 Gyr and metallicity $-$0.08 dex, while the dashed line shows solar level metallicity at the same age and dash-dot shows an isochrone of 1.3 Gyr at metallicity $-$0.08 dex for comparison.}
\end{figure}

Although cluster membership has already been established for our
target sample, it is prudent to confirm this, and in doing so also
confirm our determination of stellar parameters. To this end, we
performed a least squares fit to a set of Yale-Yonsei (Y$^2$)
\citep{YYref} isochrones against T$_{\mathrm{eff}}$ and log \emph{g}. Figure \ref{yyfit}
shows the best fit isochrone, which yields a cluster age of 1.4 Gyr,
in good agreement with photometrically determined ages in previous
studies of between 1.45 and 1.75 Gyr
\citep{nordstroem1997,anthony-twarog2009}

There is a clear outlier in Figure \ref{yyfit}, the coolest of the dwarf
stars (target No. 24), which has an anomalously high log \emph{g}. The
[Fe/H] for this star, though within the limits of the typical spread
amongst all targets, is clearly deviant from the other dwarf stars. A
review of the Fe lines used in determining the log \emph{g} value revealed
no unusual trends. The radial velocity for this target, at 2.44 km/s
from the cluster mean, is on the outer envelope of those considered
members \citep{nordstroem1997}, raising the possibility that it has
been erroneously classified as a member. We cannot dismiss the
possibility that the isochrone placement of this target is due to
binarity; \citet{nordstroem1997} marked it as a single line binary (``SB1''
in their Table 1), however we find no such indications in our spectra,
which consist of 8 epochs observed over 16 days. Given its position in
the isochrone, along with the slight extremity of its radial velocity,
we are led to reject it as a non-member based on our spectroscopic
analysis.

On examination of Table \ref{abundances}, we see two giants that
strongly deviate in [Fe/H] from the rest of the giant population,
specifically Nos. 34 and 44 at $\sim${}0.15 dex below the mean. Both of
these stars also exhibit proportionally low Ba abundances, while
star No. 44 appears to have anomalously high Nd, and enhancement in several
other elements (Mg, Al, Si, and La). This contrasts starkly with
star No. 34's relative deficiency in Nd, Gd, Mg, Na and Ti.

These two targets are squarely within the radial velocity criteria for
membership and both have high membership probabilities based on proper
motion \citep{nordstroem1997}. Both likewise do not appear anomalous on
inspection of Figure \ref{yyfit}, so it is unlikely the unusual
abundance patterns are explained by mistaken membership
assignment. Binarity for No. 34 has been suggested based on a blue
excess in photometric data \citep{mermilliod1995}, but it exhibits no
indication of binarity in its spectrum. No. 44 is not a known binary
system, however we note that the binary fraction in this cluster is
fairly high \citep{nordstroem1997}. Indeed, of the 11 targets in this
study five (Nos 11, 20, 27, 24, and 5) are considered spectroscopic
binaries, though they all clump quite tightly in Figure \ref{yyfit}.
\subsection{Metallicity}
\label{sec-4-2}

Ignoring the previously mentioned probable non-member, star No. 24,
and averaging between both dwarfs and giants as a single group, we
obtain a cluster metallicity of [Fe/H]$=$$-$0.03$\pm${}0.02. If we treat
dwarfs and giants as two separate but equally significant groups, we
obtain the slightly lower value of [Fe/H]$=$$-$0.08$\pm${}0.03.
\citet{anthony-twarog2009}, using a weighting scheme based on standard
errors, report the cluster mean [Fe/H]$=$$-$0.08$\pm${}0.02, in perfect
agreement with the second value above and only slightly poorer than
the first. A perplexing result is that the relation between group
means of giants and dwarfs, if considered separately, is reversed in
our analysis as compared to that in \citet{anthony-twarog2009}. Where we
compute [Fe/H]\(_{\mathrm{G}}\)$=$0.00$\pm${}0.03, they obtain
[Fe/H]\(_{\mathrm{G}}\)$=$$-$0.17$\pm${}0.08, and while our dwarfs have
[Fe/H]\(_{\mathrm{D}}\)$=$$-$0.16$\pm${}0.05, theirs have
[Fe/H]\(_{\mathrm{D}}\)$=$$-$0.04$\pm${}0.11, differences of 0.16 and 0.12
dex respectively in \emph{opposite} directions. It is important to note
that these mean abundances for dwarfs are actually compatible within
uncertainties, and that our value is based on only two dwarfs as
compared to their thirteen. It is possible that the discrepancy
between giants in our analyses can be attributed to differences in
stellar parameter determination as described in
\citet{anthony-twarog2009}; here we use ionisation balance, while they
use photometric methods. Referring to Table \ref{uncertainties}, had we
adopted the stellar parameters of \citet{santos2009}, the mean cluster
metalicity would increase by $\sim${}0.12 dex to [Fe/H]$=$0.04.

In spite of the differences in analysis mentioned above, and noting
the paucity of abundances in the literature for this cluster, we now
attempt to make comparisons of other element
abundances. \citet{anthony-twarog2009} were able to derive abundances of
Si and Ni for both giants and dwarfs, and additionally Ca for their
dwarfs. Between our studies for giants, Ni is compatible at close to
solar values, yet Si in their analysis is significantly enhanced
compared to ours, at [Si/Fe]\(_\mathrm{G}\)$=$0.22$\pm${}0.14, though this
could be explained by the difference in numbers of lines used (their one
line versus ten in the present analysis). For the dwarfs, both Si and
Ni are at similar levels while Ca is 0.10 dex below our mean but still
within uncertainties.  Another study, \citet{pace2008}, derived
abundances based on two dwarfs for Na, Al, Si, Ca, Ti, Cr and Ni
obtaining 0.01$\pm${}0.08, $-$0.08$\pm${}0.04, $-$0.01$\pm${}0.04,
0.04$\pm${}0.06, 0.04$\pm${}0.08, 0.01$\pm${}0.04, and $-$0.05$\pm${}0.03
respectively. With the exception of Na and Cr, which seem at odds, we
see no major differences amongst dwarf stars. Both Na and Cr for
dwarfs exhibit the largest uncertainties amongst our entire stellar
sample.
\subsection{Gadolinium}
\label{sec-4-3}

While synthesizing Nd we noticed a feature in the spectra that could
only be properly reproduced by modulating the abundance levels of the
r-process element Gadolinium. This feature also appears and is well
fit with our atomic data in the solar spectrum.  To our knowledge,
there exist no abundance measurements of Gd in open clusters in
the literature. \citet{sneden1983}, using several Gd features at 3549\AA{}
and 3768\AA{} in the spectrum of HD 122563, obtained a similarly
deficient abundance of [Gd/Fe]$=$$-$0.50, and \citet{denhartog2006} addressed
the lack of measurements by improving abundance measurements of the Sun
and deriving abundances for three metal poor giants obtaining values
of $-$0.14, $-$0.42, and $-$1.08 respectively. Unfortunately, as these are
not known cluster members and were selected for their low metal
abundances, it is difficult to place those results in the context of
the current study. Our measurements are based on a single line at
4463\AA{}, which is evident in the solar spectrum and likewise can be
fit in the NGC 3680 giant spectra. 

\begin{figure}
\centering
\includegraphics[width=\linewidth]{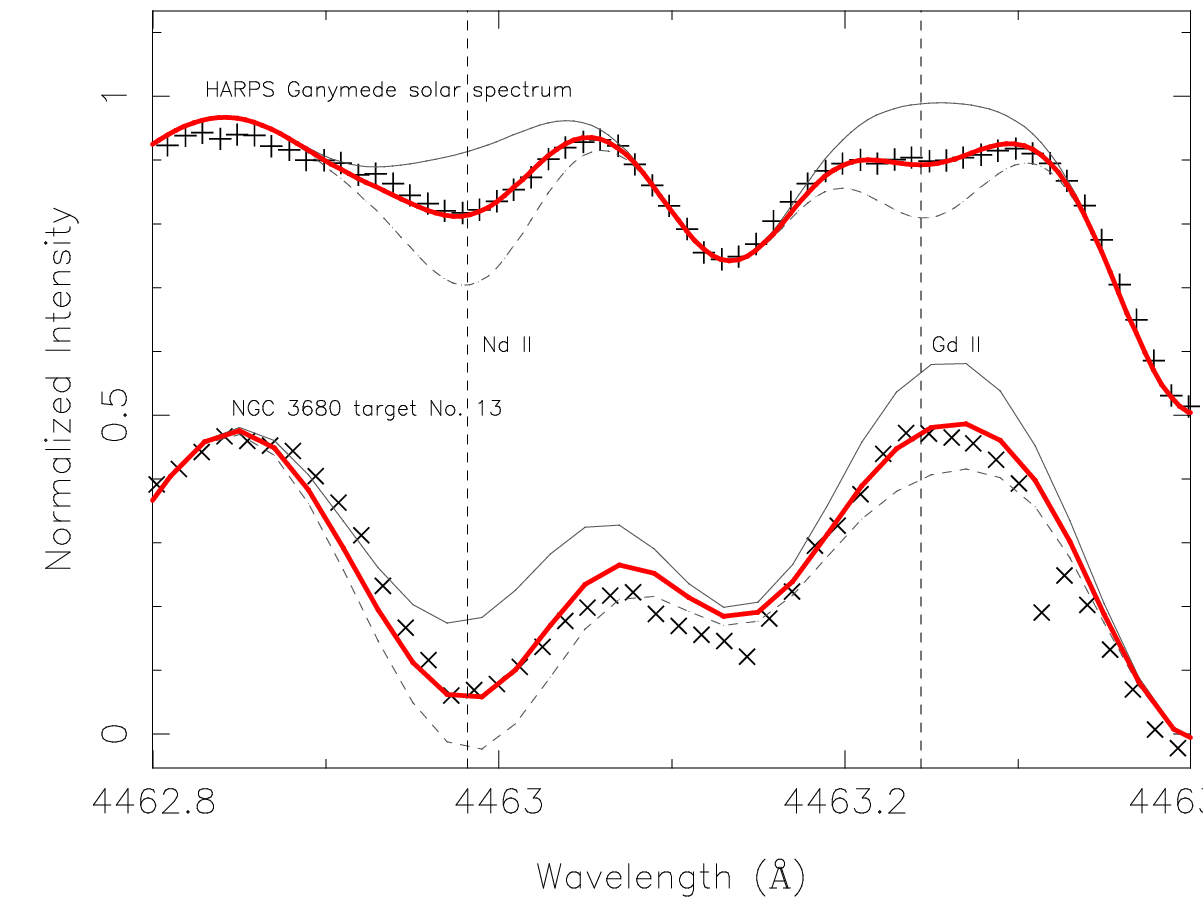}
\caption{\label{gdsynth}Nd and Gd synthesis for the HARPS Ganymede and one of our target giant star spectra (plus and x symbols respectively). The thick red lines indicate our accepted fits, while the thin solid and thin dashed lines give an idea of the sensitivity, representing $-$0.4/$+$0.5 dex for Nd and $-$1.3/$+$0.2 for Gd in star No. 13, and $-$1.6/$+$0.4 for Gd in the sun}
\end{figure}

Figure \ref{gdsynth} shows synthesis results for Nd and Gd for both the
Ganymede solar spectrum and a representative giant spectrum from our
analysis; note the difference in line shape between the giant star and
the Sun, which is likely caused by broader profiles of the Fe features
on either side. Our computed solar abundance of Gd, using line data
obtained from VALD, is 2.77 which is well above that computed by
\citet{denhartog2006} of 1.11. Effects such as hyper-fine structure or
isotopic ratios may effect the computations significantly and were not
included in our synthesis, as we are not aware of any analysis
specifically involving the line at 4463\AA{}. With the cluster giants at
near solar metallicity, the robustness of our differential values
seems enhanced; however effects due to temperature and log \emph{g}
(i.e., comparing giants and dwarfs) may play a significant role in the
shaping the Gd profile, making such comparison difficult.

Eu synthesis proved to be too difficult for this cluster so Gd is the
only pure r-process element we were able to measure. Gd may act as a
proxy for the trends of other rare earth abundances in this cluster;
however given the lack of specific knowledge of this species, in
addition to our measured solar value well beyond literature solar
values, our reported abundance must be taken with caution.

\subsection{Na enhancement in giants vs. dwarfs}
\label{sec-4-4}

There is a clear divide between giants and dwarfs in terms of [Na/Fe]
abundances, of order 0.25 dex, as seen in Table
\ref{abundances}. Assuming that the stars in this cluster formed out of
a well mixed common natal environment (e.g. \citealt{fbh2002}), it is
possible that the initial abundances of Na in RGB phase stars are
modified and enhanced via the Ne-Na cycle, after which deep convection
brings the newly formed Na to the photosphere \citep{salaris2002}. It
is unclear whether the temperatures required for the Ne-Na cycle are
reached in these stars, and thus a primordial explanation may be
required. The commonly observed O-Na anti-correlation in Galactic
Globular Clusters (GCs) may be related to this action, as the CNO
cycle is thought to require the same temperatures/depths as the Ne-Na
cycle, and is responsible for the depletion of O. Again, whether this
observation is explained by stellar evolution or primordial abundance
variations is still an open question. It is interesting to note,
though, that the same anti-correlation has not been observed in OCs in the
Galaxy \citep{desilva2009}.

It is also possible that the abundance differences between dwarfs and
giants are the result of non-LTE effects and no intrinsic differences
exist. Na is particularly sensitive to non-LTE effects, though the
most notable corrections are required at extremely low metallicities
\citep{andrievsky2007,baumueller1998}. Even after computing Na
abundances using non-LTE models, \citet{andrievsky2007} observe a Na
enhancement in giants, also present in their LTE analysis, which they
attribute to a mixing mechanism as above.
\subsection{The s-process element abundances}
\label{sec-4-5}

\begin{figure}
\centering
\includegraphics[width=\linewidth]{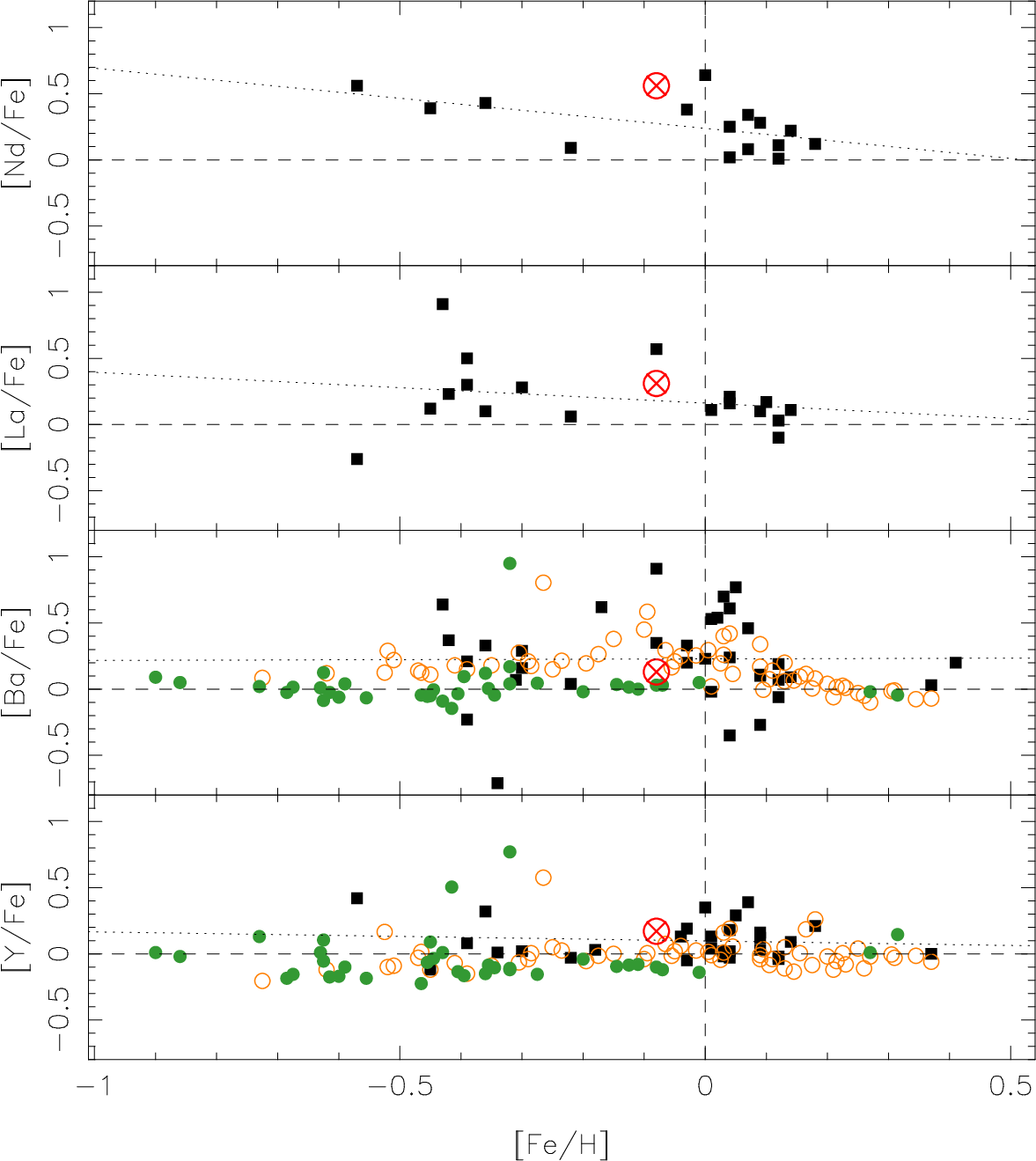}
\caption{\label{heavies}[Fe/H] vs. heavy metal [X/Fe] abundances in a selection of open clusters and Galactic field stars. Filled squares represent cluster means as described in the text, while open and filled circles represent thin and thick disk field stars, respectively. NGC 3680 is shown by the encircled x symbol. Vertical and horizontal dashed lines mark solar abundance levels. The dotted line in each panel is a linear regression to the cluster abundance data, for illustrative purposes.}
\end{figure}

Due to the lack of a range of s-process abundance measurements for NGC
3680 in the literature, we take this opportunity to discuss abundance
trends for the elements Y, Ba, La, and Nd specifically. Figure
\ref{heavies} places the heavy element abundances of NGC 3680 within the
context of other OCs and Galactic field stars where possible. The
thick and thin disk samples were taken from \citet{bensby2005} while the
cluster sample was compiled from references within \citet{carrera2011},
using the mean abundance value from available high resolution studies
for a given cluster. NGC 3680 exhibits higher than average La and Nd
abundances, although several other clusters show more enhanced
abundances for these two species. IC 4756 and NGC 2420 exhibit higher
Nd abundances while NGC 2141 and Tombaugh 2 are more enhanced in
La. NGC 2420 has both the lowest La and highest Y in the sample,
and is additionally the most metal-poor cluster considered here.

The two heavy s-process elements, La and Nd, appear to be more
enhanced with respect to Fe as metallicity decreases amongst the
cluster population. This trend is broadly consistent with that
predicted in AGB stars due the the increase in neutron exposure with a
decrease of the iron seed nuclei (e.g. see \citealt{busso2001}). Ba does
not seem to exhibit such a trend, though the cluster abundance scatter
is quite high. \citet{d'orazi2009} also find no correlation in [Ba/Fe]
with metallicity amongst clusters, but notice a significant
anti-correlation with age, which they argue is evidence of a greater
extra-mixing efficiency for lower mass stars producing the $^{\mathrm{13}}$C
neutron source. NGC 3680, in our analysis, falls roughly 0.06 dex
below the group of intermediate age clusters shown in Figure 2 of
\citet{d'orazi2009} roughly 0.06 dex below the group of intermediate
ages, but still quite clearly within the uncertainties and thus does
not challenge the age-Ba anti-correlation observed.

\begin{figure}
\centering
\includegraphics[width=\linewidth]{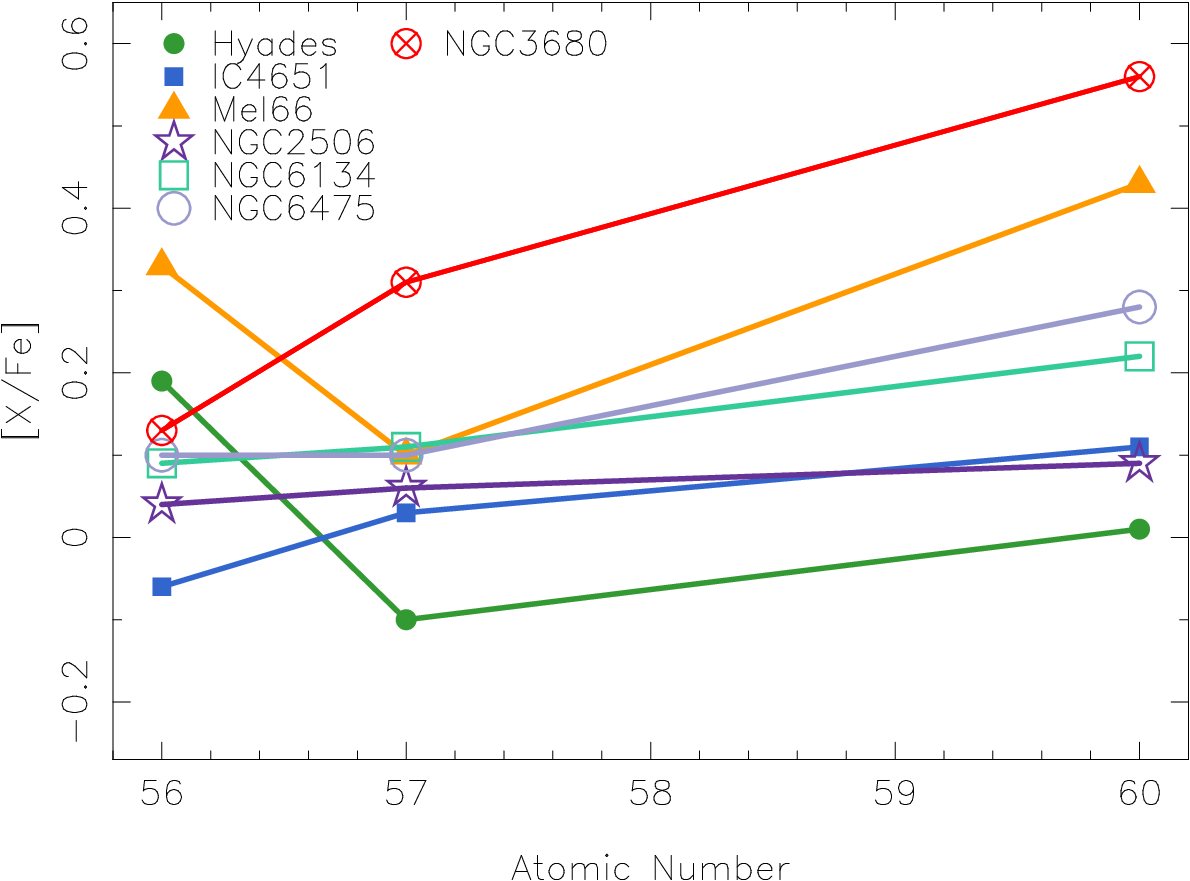}
\caption{\label{heavy_s}Heavy element trends with respect to atomic number for a subset of the cluster sample shown in Figure \ref{heavies}, NGC 3680 indicated by the red line and encircled x symbol.}
\end{figure}

The heavy s-process elements Ba, La, and Nd in the giant population of
NGC 3680 have increasing abundances with increasing atomic number, by
roughly 0.2 dex at each step. Other OCs exhibit a similar
monotonically increasing trend, as seen in Figure \ref{heavy_s}, which
plots atomic number vs. mean cluster abundance for a sample of
clusters from Figure \ref{heavies} which had measurements for all three
elements, though none with a slope as consistently high as
NGC 3680. Since the s-process alone from models of AGB stars yields a
clear peak at Ba resulting in lower abundances for the other heavy
s-process elements \citep{busso1999} across a range of metallicity and
mass, another process is needed to explain the observed
enhancements. The most likely such process is the rapid neutron
capture (r-) process, which quickly builds very large nuclei that
decay over time to the stable isotopes beyond atomic numbers of
around 55. This process is thought to require high neutron densities,
likely only available in the shocks of core-collapse supernovae. The
Ba peak from AGB models is observed in solar photospheric and
meteoritic data (e.g. see \citealt{asplund2009}), implying a higher ratio
of high-mass stars (those fated for the SN II) in the cluster's
progenitor population, as compared to that of the Sun. Other clusters
exemplify the Ba-peak (e.g. see Figure 2 of \citealt{desilva2009b}),
indicating a range of formation scenarios that can be explored using
abundance trends within the s-process elements.

The sample in Figure \ref{heavy_s} has a limited range in [Fe/H], owing
in part to the limited availability of literature abundances for these
heavy elements. Both the r- and s-processes have a dependence on
metallicity due to the seeding nature of the Fe nuclei; more data are
clearly needed at a range of metallicities lower than [Fe/H] $\approx${}$-$0.5 to
explore this effect in more detail.

\section{Conclusions}
\label{sec-5}

We have performed a detailed spectroscopic abundance analysis on the
intermediate aged open cluster NGC 3680, obtaining differential
abundances of Na, Mg, Si, Ca, Ti, Cr, Ni, Zn, and Y for 8 giant and 3
dwarf stars, and Al, Ba, La, Nd, and Gd for the 8 giant stars. We find
a combined metallicity of [Fe/H]$=$$-$0.03 dex, and for the giants in this
study [Fe/H]$=$$-$0.08 dex in good agreement with literature values. Based
on our analysis, we propose that the membership assignment of a single
star (No. 24 in this study) be rejected both due to its position on
the CMD and its anomalous metallicity.

The serendipitous measurement of Gd in this cluster represents the
only measurement of a pure r-process element for this cluster, and
likewise the only measurement of Gd for an open cluster that we are
aware of. This may have important implications with respect to the
progenitor population of NGC 3680, however due to measurement
uncertainties it is difficult to comment on its value.

A significant divide in [Na/Fe] abundances between dwarfs and giants
is observed. This may be due to non-LTE effects which are difficult to
quantify, but there is also the possibility that convection in the
giants is bringing fresh Na to the photosphere, thus enhancing the
[Na/Fe] abundance. If that is the case, this action may be related to
the O-Na anti-correlation observed in Galactic GCs, but not OCs, though
the mechanism responsible for this is still not well understood.

An important result of this analysis is the measurement of a range of
s-process elements (Y, Ba, La, Nd), for which measurements in open
clusters remain limited in the literature. Comparing these abundances
with available open cluster data, we find that NGC 3680 fits within
the typical spread, with La and Nd having above average values. The Ba
abundance from our analysis is consistent within uncertainties, though
somewhat low for the cluster's age, in the context of the recent
finding by \citet{d'orazi2009} of an anti-correlation of [Ba/Fe] with
age.

We observe a step-like enhancement within the heavy s-process peak
elements in this and other clusters. We argue that this observed trend
indicates a substantial amount of r-processing contributed to the
generation of La and Nd, likely in core-collapse supernovae in the
progenitor populations of these clusters.
\section*{Acknowledgments}
\label{sec-6}

The authors would like to acknowledge Yuan-Sen Ting for providing the
table of abundance data compiled from the references within
\citet{carrera2011}. We thank Valentina D'Orazi for helpful advice on
spectrum synthesis. Finally, we would like to thank our anonymous
referee for helpful suggestions in strengthening this work.

\bibliographystyle{mn2e}
\bibliography{ngc3680_paper}
\end{document}